# Characterization of Electrical Impedance Tomography System

Vaishali Sharma, Mayank Goswami[#]
*Divyadrishti Imaging Laboratory,*
*Department of Physics,*
Indian Institute of Technology Roorkee, Roorkee, Uttarakhand-247667, India.

[#]Mayank.goswami@ph.iitr.ac.in

## Abstract
Electrical Impedance Tomography (EIT) can be cost-effective, portable, non-invasive imaging technique. It has pre-clinical and a few of them already proven industrial applications. This technique can only recover images of low spatial and contrast resolution, partially due to existing physical models. The capability of discriminating between Impedance profiles in recovery is around 73%. However, similar to other modalities, EITs' performance depends on the hardware and recovery algorithm design and operating parameters. This work presents an empirically obtained mutual relation between the hardware design-related six independent variables, namely: (a) molarity of the coupling media, (b) scanning duration, (c) Size of the object, and (d) parameters defining the size of the scanning assembly (No. of electrodes, area of the vessel, and percentage periphery covered by the electrodes), affecting its performance. The expression predicts that the error can be kept under a 10% value in a worst-case scenario if these six parameters are kept under a given range. The root mean square error between the experiment values and predicted values from the presented equation is 1.4053. It is shown that time significantly affects the recovery process compared to other optimizing parameters. It is also shown that the accepted molarity value for the presented system is 2M.

## 1. Introduction
### 1.1 Classification of system designs

Electrical Impedance Tomography (EIT) is a non-invasive imaging technique that utilizes relative changes in the pattern of transmitted current (10-100kHz) of the order of milli-ampere under a region of interest (ROI) [1], [2]. The three-dimensional region (cross-sectional area having a thickness equal to the length of electrodes) that is covered by electrodes is referred as ROI. The reconstructed image represents averaged (over the thickness) two-dimensional impedance distribution of ROI. For a typical EIT system, the ROI is defined as a region inside a container onto which a set of electrodes are mounted, and it is filled with a conductive fluid (for example, saline water) as a part of material distribution along with another phase of material [3]–[5]. This EIT configuration having a multiphase inner profile distribution is referred to here as a Type-I EIT system. Type-II EIT system, however, has no container. Electrodes are mounted on a belt/holder or directly over the object using a primary couplant [6], [7]. Type -I EIT may or may not need any couplant between the container and electrode surface. System images of these two types are shown in section 3.

ROI with overall high impedance inner distribution would allow a relatively low depth of penetration as compared to low impedance inner distribution, thus limiting EITs' probing capabilities, respectively. This fact basically indicates that the Type -II EIT system will always generate a sub-surface non-invasive image. Even for sub-surface non-invasive imaging to achieve an acceptable depth of penetration, system design requirements need optimization [8]–[11]. A solid copper cylinder having a crack inside its deep core would require Type – II EIT system. Type-I (when this cylinder is placed inside a beaker filled with saline solution), similar to Type-II systems, would be able to reveal only its surface topography.

A customized commercial Type-I EIT system would only reveal the location of bubbles and slurry flowing inside a polyethylene terephthalate pipe. It may not be able to probe and reconstruct if slurry material is entered inside any of those bubbles (having thick boundaries). Taking pipe (with the objective to probe multiphase flow imaging) as an example, EIT can be defined as a fully-fledged Computerized Tomography (CT) system. EIT systems are used to probe industrial applications such as multiphase flow imaging, crack analysis, blockage of pipelines, lifespan, and moisture of a wall [12]–[22].

## 1.2 Resolution and associated parameters

Mathematically, the EIT modeling is an inverse solution process that may be ill-conditioned [23]–[26]. The finite element method is used to discretize the Laplace equation with an adaptive grid. The effect of grid size (number of pixels used to discretize the ROI) on spatial resolution saturates after a certain value [27].

The possible recoverable details (for example, size, shape, etc.) of discretely distributed material (for example, bubbles), i.e., the image resolution, may improve by optimizing the EIT system design and operating parameters. Electrode size, width, input signal frequency, and the number of electrodes are the only factors shown affecting the resolution [28]–[33]. Studies have shown that the best EIT results are obtained when 80-90% of the ROI is covered by electrodes[34]–[38]. We note that ~100% coverage will require insulation between successive electrodes. Existing EIT systems (reported in the literature and commercially available) cover a wide spectrum of input signal frequencies ranging between 2 kHz to 10 MHz, according to the specific application[39]–[44]. One of the optimized and preferred, the "Howland current source," limits to go beyond 100 kHz to avoid instability in current injection [45]–[50].

There are many current injections (the way a certain set of electrodes are excited) circuit configuration, namely: (a) adjacent, (b) parallel electrodes, (c) skipping-m electrodes, (d) polar (angle specific electrode selection) injection and (e) measurement methods are proposed for better signal to noise ratio (SNR) [51]–[57]. Single current source and multiple current sources can be used for electrode excitation. Skipping-m number of electrodes strategy excites two non-adjacent electrodes leaving m number of non-excited electrodes in between. It measures the output from the rest of the electrodes. The skip-0 pattern shows the use of adjacent electrodes for excitation. The opposite electrode strategy is also used in which electrodes that are diagonally placed are used for excitation so that signal can cover the whole diameter. Studies showed that there is no way to determine one best current injection method as every method provides an improved image in a different sense [58].

To the best of our knowledge, for EIT design optimization: (a) injection circuit configuration, (b) peripherical coverage by an electrode, (c) frequency responses, and (d) inverse algorithm selection are considered only. A typical Type-I EIT laboratory prototype system uses saline solution (mimicking intra-cavity fluid for clinical applications or fluid in industrial multiphase flow channel) filled in a container (mimicking the periphery of nonconductive pipe, human body, the first surface of a wall, etc.,) [59]–[62]. This saline solution's molarity is an important design parameter that may affect the resolution of the EIT reconstruction. Externally induced ionic flow due to electrolysis may affect the concentration of a saline solution in the ROI. The extent of this effect depends on how long the EIT measurement is performed. Thus, the experiments' duration may also be one of the design optimization factors. We note that state of art EIT current circuit design (Howland current source) has limitation for upper current values that limits the depth of penetration of electromagnetic wave. It may have no interaction with a discretely distributed patch of material of a certain size for a given current value EM wave. The interaction, however, may improve if electrodes are placed nearby to that particular patch, however, it creates an adaptive electrode placement configuration [63]. Thus, the ratio of system and object size may also be another factor.

## 1.1 Motivation:

Only a few of the typical EIT system design parameters are experimentally studied with the objective of optimizing the performance (in terms of resolution). The individual effect of number, size, type of electrodes, and injection current frequency is already reported.

We have incorporated the effect of the molarity of a saline solution, the size of the object relative to the size of the system, and the scanning duration. Most importantly, *mutual interdependencies* of already studied parameters: number, size, and type of electrodes and newly added factors: molarity of a saline solution, size of the object relative to the size of the system, and scanning duration are presented in this work now.

## 2. Material and Method:

A description of the theoretical modeling that generates a system of equations pertaining to EIT is briefly discussed. Solution approaches are described briefly afterwards. EIT design and experimental methodology used to test the mutual relation between design optimization parameters are described at the end of this section.

### 2.1 Mathematical Model:

$$\nabla \cdot (\sigma \nabla \phi) = 0 \qquad (1)$$

The equation to be solved in EIT is a second-order elliptical differential equation.

Boundary current is given by $j = -J \cdot n = \sigma \nabla \varphi$, where n is the unit normal vector to the ROI. The solution of Eq. (1) is obtained numerically using appropriate boundary conditions (Dirichlet and Neumann). This equation is not linear in nature, and after solving, it provides information about the conductivity/resistivity of ROI. Complete information about the nature of the problem (reported in previous studies as well) is shown in S1

### 2.2 Experimental Setup and Methodology:

Type-I EIT systems with different container sizes, different numbers, and sizes of electrodes are constructed to test which combination would provide better imaging results using set of test objects. Electronics is developed in-house. Data acquisition, signal processing, and image reconstruction codes (EIDORS v3.10) are developed/integrated with a single pipeline. Each setup is fully automated via a single push of button codes on a single programming platform using MATLAB®.

AC current is used to excite the electrodes which is generated by a voltage-controlled current source (VCCS) regulated by an AC voltage. For VCCS modified Howland circuit has been used in Figure 1(a). Two 741 operational amplifier ICs, along with some resistors, have been used to construct it. The generated sinusoidal AC current is used to excite the electrodes using two high-speed complementary metal oxide semiconductors (CMOS). CD74HC4067 ICs have been used for switching. These two CMOS are used to switch the current between electrodes. Simultaneously two other CMOS are used to read the voltage between the remaining electrodes. Output from these two CMOS is injected in a differential voltmeter[64] (figure 1b) and then amplified by an amplifier, and finally read by Arduino Mega 2560. A complete diagram (reported in previous studies as well) of the system is shown in S2. The schematic of the switching module along with its PCB form is shown in Figure 1.

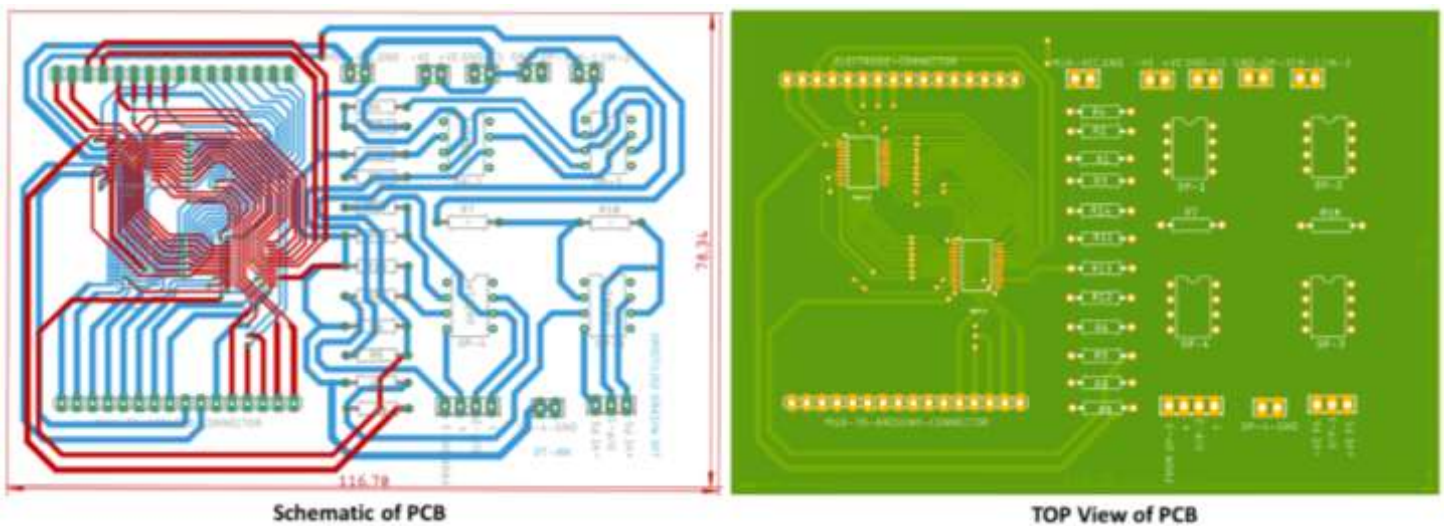

**Figure** 1. Circuit Diagram of EIT Switching Module

This version is submitted to arXiv on 14/04/2023

For experiments, a number of metallic electrodes are distributed by covering the region of interest. Electrodes are subjected to an injection of a constant ac current at a frequency of 50 kHz and 4 $V_{RMS}$. At this combination, the circuit is facilitating a maximum of one $mA$ current. The response of ROI depends on the frequency of the signal, as discussed above. We have used the adjacent electrodes strategy to excite the electrodes and data collection. After applying the current, the potential difference is measured from the boundary electrodes. This process is repeated sequentially for every possible combination of the electrodes. Figure (2) depicts the schematic images (a container with EIT electrodes and corresponding cross-sectional view) and measurement scheme for a configuration with sixteen electrodes. Firstly, electrodes 1 and 2 are excited, and the potential difference is measured between other consecutive electrodes starting from electrode no. 3 till 16. Potential difference at electrodes 1 and 2 is not included in the process of reconstruction as including it increased the noise in our case. Figure 3 shows an in-house developed EIT experimental setup along with the embedded electronics (switching module). The top view of the EIT vessel/container is shown at the top right corner of the image. It is converted into a compact PCB-based setup.

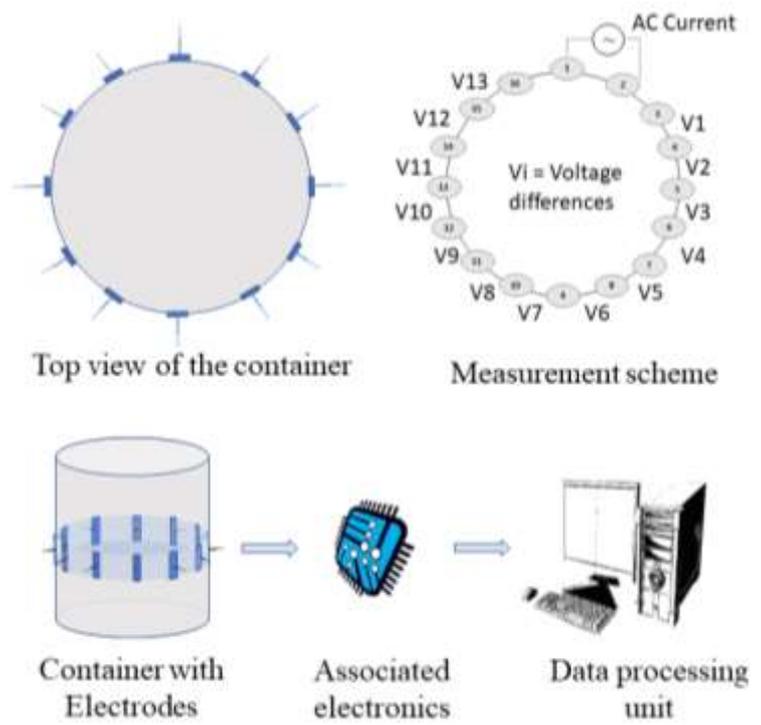

**Figure** 2. Schematic image and measurement strategy of EIT system

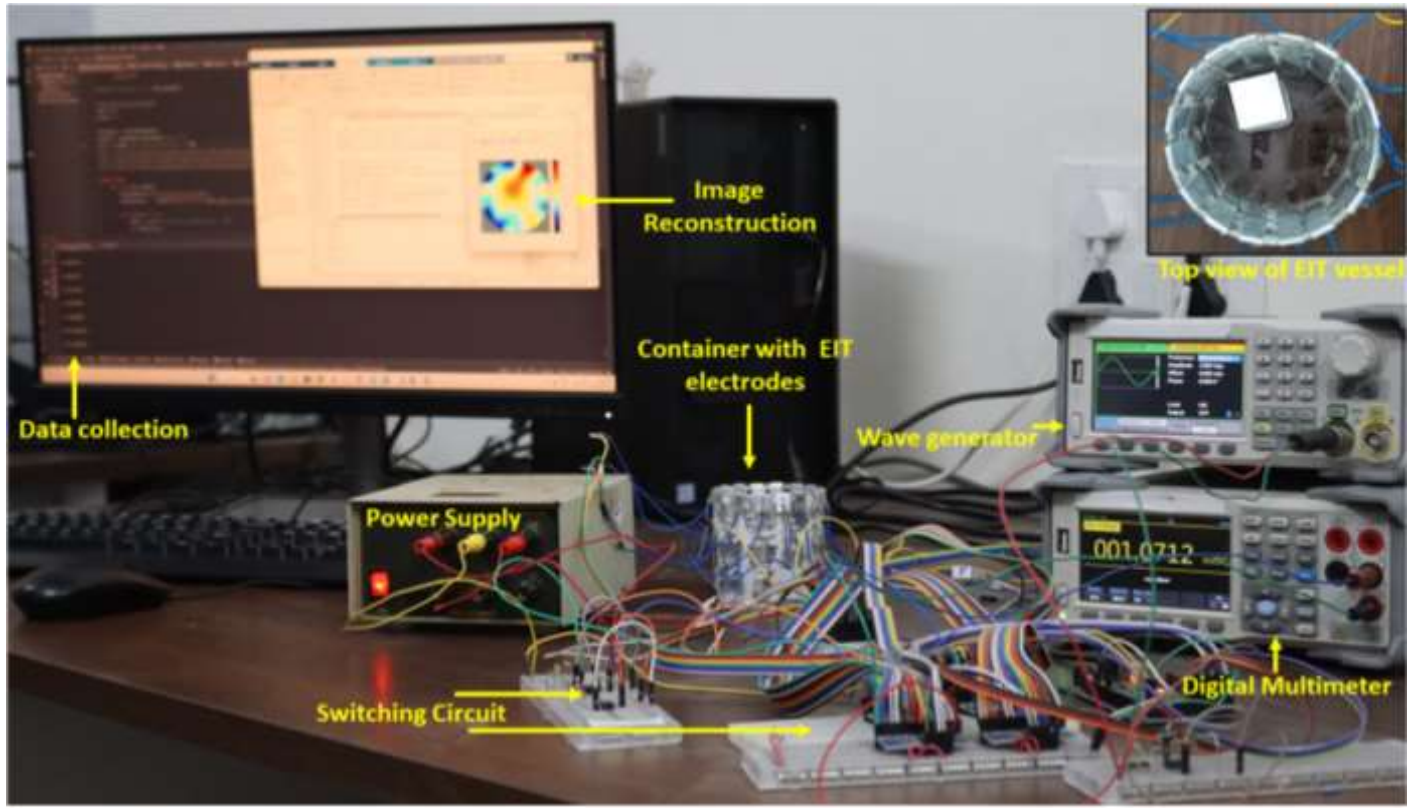

**Figure** 3. Perspective view of real-world photo of Experimental Setup.

This version is submitted to arXiv on 14/04/2023

## 2.3 Variation of design parameters:

**Preparation of the molar solution**: Molar solutions of different molarities are prepared according to the formula described below:

$$Molarity = \frac{Mass(solute)}{(Molucular\ Mass)*Volume}$$

We determined the molarity and prepared solutions in accordance with the results for different masses of solute (NaCl).

**Sample preparation:** Ultimaker Extended 2+ is used to print three-dimensional samples. To start, Solidworks software is used to create Computer-Aided Design (CAD) files. Each sample has a cuboidal shape and a 5.5 cm height, according to the CAD files for the samples. The samples' width and length are adjusted by 0.2 cm intervals between 0.1 cm and 1.9 cm. The smallest phantom has dimensions of 0.1 cm(W) x 0.1 cm(B) x 5.5 cm(H), while the largest size sample is manufactured and has dimensions of 2.5 cm(W) x 2.5 cm(B) x 5.5 cm(H). The parameters are kept at 150 mm, 45 mm/s, 80 percent, and grid, respectively, for layer height, fill density, and fill pattern. Following that, Ultimaker Cura 4.13.1 software is used to transform the CAD files into .gcode files. At 45°C and 235°C, respectively, the bed temperature and nozzle temperature are set. Ploy-Lactic Acid (PLA) is the ink used to print the samples, and it has a 0.2 mm nozzle.

## 2.4 Reconstruction

EIT images are reconstructed using EIDORS, a MATLAB® based software [65]. ROI is discretized with 3530 nodes using adaptive Delaunay triangulation. The code creates a container of a given dimension in the cyber world to specify the boundaries. The code has its own terminologies to select the inputs. No measurement current, no rotation measured, sixteen electrodes, adjacent injection pattern, and 1mA current options are selected for stimulation patterns. Measured real world, EIT data is fed to an inbuilt inverse solver that generates a final reconstructed 2D image. The code requires information of the common FEM model along with measured data.

A typical reconstruction algorithm expects projection data from all views together. Projection data is formed by measured data in the presence of an object and the absence of an object as background. In all the reconstructions shown in this work, the background is measured when only saline water is filled inside the container. Gauss-Newton method is selected for the inverse solver.

## 3. Results and discussion:

### 3.1 EIT system characterization

Type-I EIT systems are used throughout the remaining text. All experiments are repeated three times to plot the error bars. To see the effects of vibration and temperature on EIT systems, experiments using a magnetic stirrer and rocker are performed (Movie1). After analyzing the data from multiple experiments, it is found that these two parameters do not affect significantly (have at maximum 0.88 and 0.63 standard deviation) the EIT performance. These outcomes could be attributed to the following reasons (i) high frequency of the AC current leads to a more rapid alteration of the electric field, resulting in a greater force acting on the ions, causing them to move with increased speed. As a consequence, this can lead to ion motion and enhanced mixing in the solution (ii) conductivity of saline solution increases with temperature as well, which leads to moving ions faster in the solution this motion of ion. Impact of both the parameters through error bar plots is shown in figure4 (original and scale-adjusted plots). Following the conclusion, all subsequent experiments for further investigation are carried out under controlled conditions of room temperature and minimal vibrations using plastic containers (Polypropylene Plastic).

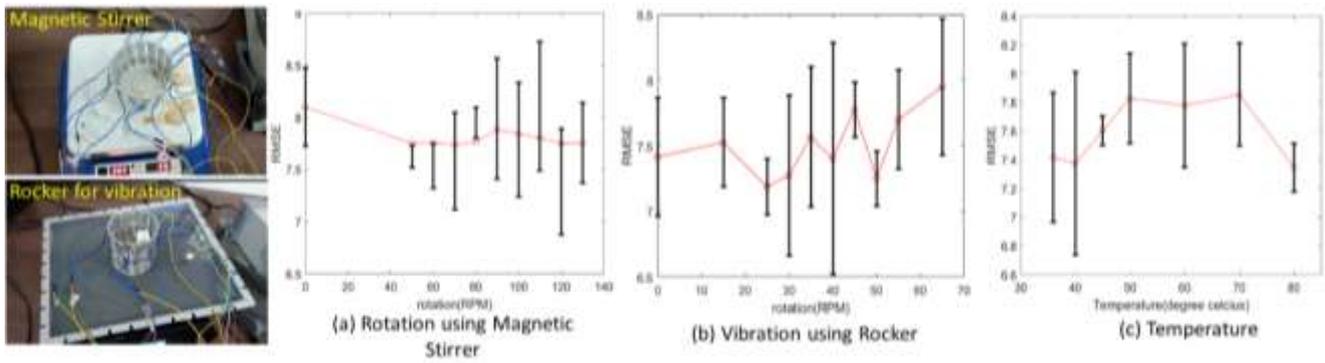

**Figure** 4. Reconstruction quality and variation of temperature and vibration; Error in reconstruction when EIT with the sample inside is stationary; however: a) saline water is being rotated, b) is placed on a vibrating platform, and c) saline water having different temperatures is used. The experiments are recorded and shown in Movie1. The variation in quality is negligible, having maximum 0.61, 0.88, 0.63 standard deviations, respectively.

### 3.1.1 High and low impedance discrimination

Two objects (object 1 and object 2), with different impedance values, are used together and separately as the primary material to create three different specimens of heterogeneous inner profile arrangements merged in saline media as secondary material. The real-world photos of these three specimens are shown in Figures. 5(a), 5(c) and 5(e). Object 1, with a 3.3 cm diameter aluminum cylinder, has a 1 cm diameter air column. Object 2 has an outer part made of a rubber cylinder that consists of three metal rods inside. The wireframe replicas with the corresponding cross-sectional view of both objects are shown in Figures. 5(g, h) and 5(i, j).

Sixteen electrode system is constructed. These electrodes with a 1.7 cm electrode width are distributed over a container having a diameter of 21.7 cm. The radially converging lines (Fig. 5(a), 5(c), and 5(e)) (highlighted using a red arrow) in the bottom are just markings made by pen to show the position of vertical electrodes. The reconstructed images of specimens are shown in Figs. 5(b), 5(d) and 5(f) respectively. A fake color scheme (parula in MATLAB®) is used. These images depict the low-impedance object 1 (aluminum) in dark blue and the high-impedance object 2 in red. The light blue color depicts the presence of saline solution. The reconstructed images fail to estimate exact shapes. The reconstructed profiles of object 1 and object 2 do not have sharp boundaries but are diffused into a lighter color. It fails to recover the column (filled with water) inside the aluminum cylinder of object 1. It also could not recover the three metal rods inside object 2. In certain random locations near the periphery of the container, random shapes and colored artifacts are also visible (highlighted by an arrow marker of purple color). Several other research groups have reported the existence of the same artifacts. All of them have used EIDORS. Figure 5(b) shows that EIT is able to discriminate two different objects with their relative size and locations with respect to the periphery of the container. The percentage difference between true impedance values is around 99.99% for sample 1 shown in Figure 5(a). The percentage difference between recovered grey/index values (shown in Figure 5(b)) is 73.29%. Qualitatively, one can also observe that EIT is able to relatively discriminate the high and low impedance inner profile distribution.

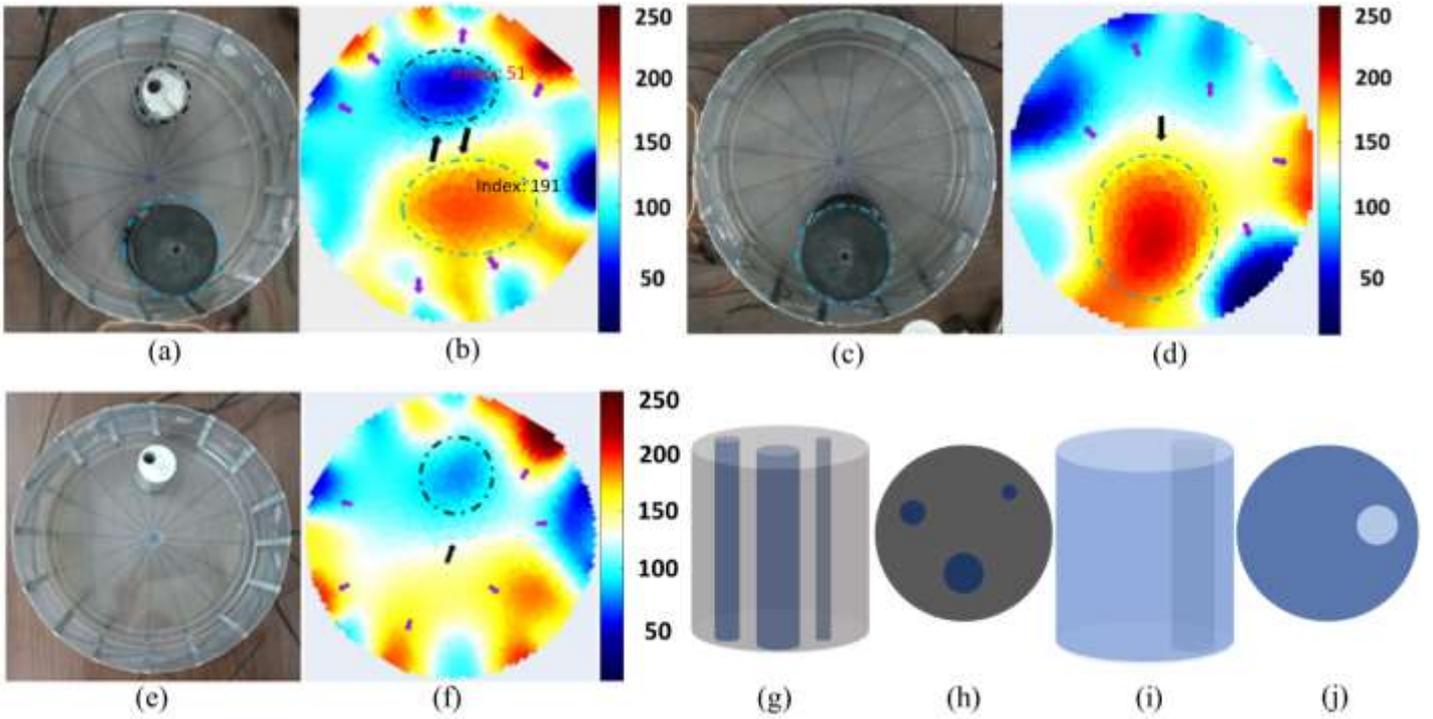

**Figure** 5. EIT resolves the High and low impedance structure; a) shows the sample having high and low impedance distribution, b) recovered by EIT, c) high impedance inner profile recovered in d), similarly e) shows that low impedance inner profile is recovered in f); g)-j) shows a graphical representation of the sample.

### 3.1.2 Inherent error in reconstruction

Specimen 4 is created by submerging a 3D printed square column object (0.7 cm edge), made of PLA inside a 5.4 cm-diameter container (filled with saline solution) at the center location (figs. 6(a)). Its cyber replica is created and shown in 6(b). The fake color scheme is chosen according to the relative impedance values of PLA and saline water. The cyber replica is used to create noise-free projection data by solving forward problems. This noise-free synthetic projection data is used to reconstruct back the profile, as shown in 6(c). The estimated inherent error in terms of root mean square (RMSE) value is 6.5.

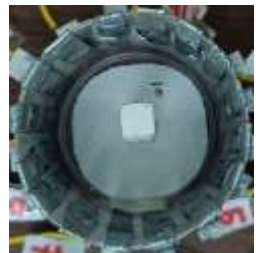
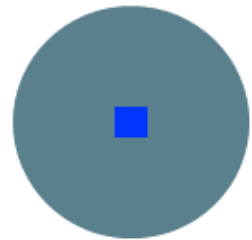
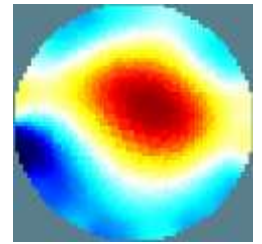

Fig. 6(a) Real photo (Top view)    Fig. 6(b) Phantom    Fig. 6(c) Reconstruction of Phantom

### 3.2. Design parameter optimization

Several similar objects with variable sizes, shown in Figure 7 are scanned. A flat, fragile base, if necessary, is provided to support these samples so that they can stand inside the container. This exercise is performed to study the effect of the size of objects to design optimization parameters (one at a time). The height of the scanning is kept in the middle of each sample.

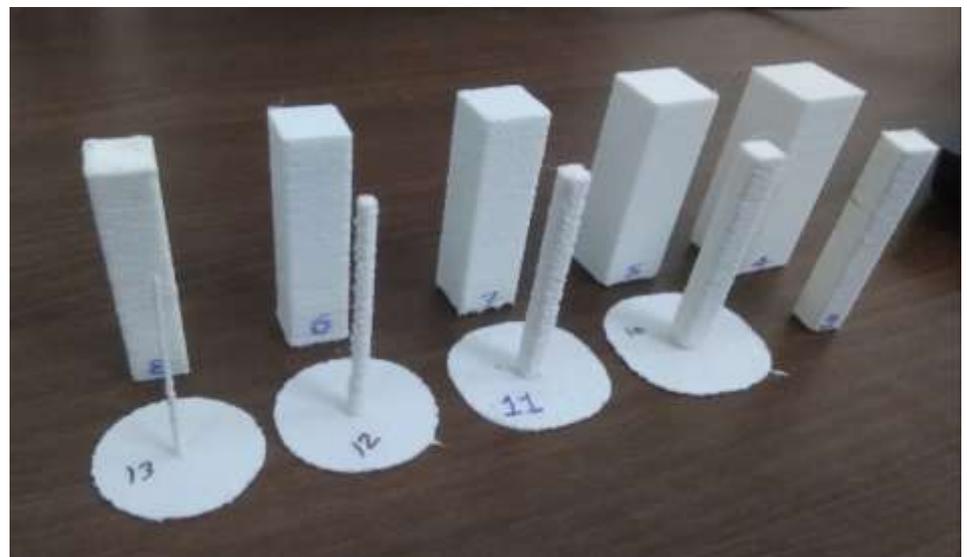

**Figure** 7. Ten 3D printed PLA Objects with variable size.

Experiments are performed using different EIT vessels by changing only one design optimization parameter at a time.

### 3.2.1 Molarity of coupling media (saline solution):

Numerous experiments have been performed to see the effects of molarity (M) on the recovery process. Solutions with different molarities (0.4M to 2M) are used in a vessel with sixteen electrodes and 5.3 cm diameter. Reconstructed images (for each variation in M) are used to find the root mean square error (RMSE) using cyber phantom (discussed in section 3.1). In the first case, RMSEs are found for different PLA objects by keeping the scanning duration constant (20 min). In another instance, RMSEs are found for different scanning duration, maintaining PLA size (0.20 cm diameter) constant. Figures 8(a) and 8(b) demonstrate how the saline solution's molarity affects image reconstruction. Experimentally it is observed that when the molarity of the solution increases, root mean square error (RMSE) between the phantom and the reconstructed image first decreases up to a certain range (~ 0.5-0.6M) and then increases. The optimal range of molarity for this EIT (having other design parameters fixed) would be between 0.5 to 0.6. Each experiment is finished within 10 minutes to ensure electrolysis does not affect the concentration of dissolved ions.

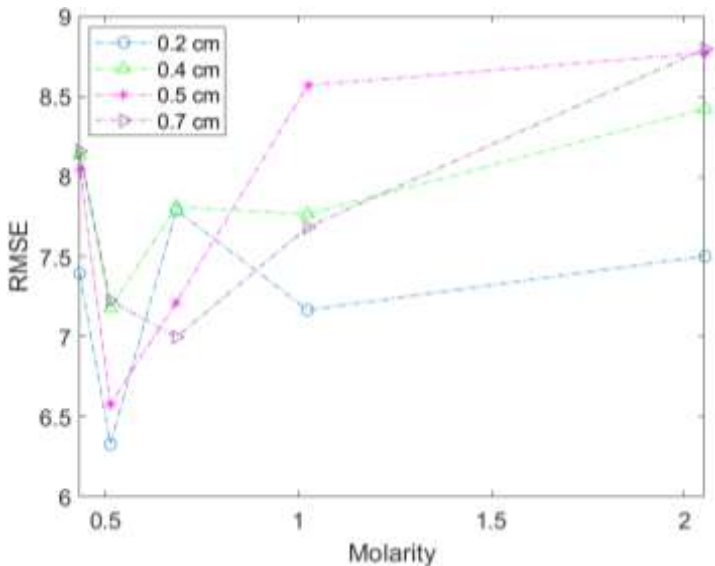
Figure 8(a)

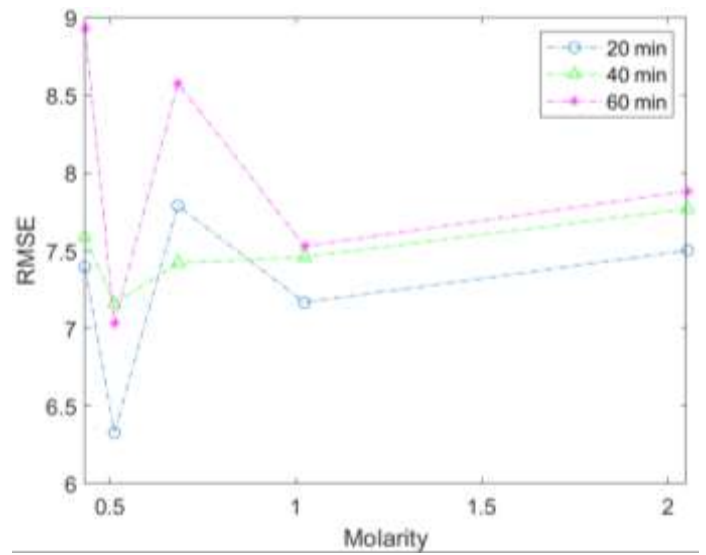
Figure 8(b)

**Figure 8**. Effect of Molarity of the saline solution on reconstruction quality. The optimal value of molarity exists for the setup used in experiments.

Table 1: Percentage change in RMSE

| Physical quantity | Percentage change in physical quantity | Max percentage change in RMSE | Min percentage change in RMSE |
|---|---|---|---|
| Molarity (M) | 1% | 1% | 0.4% |
| Time (Min) | 20% | 1.6% | 0.4% |

### 3.2.2. Scanning Duration

To carry out this study, experiments for different scanning duration along with different molarity are performed. The effect of scanning duration (spent to carry out a single measurement) on reconstruction is studied using an EIT system consisting of a container (diameter of 5.3 cm) filled with saline solution fitted with sixteen electrodes. An object having a diameter of 0.2 cm for this study. However, object sizes are also varied. Multiple experiments are performed every time the scanning duration (T) is varied while ensuring that other parameters are kept the same at the very beginning of each experiment. The effect of scanning duration on reconstruction is shown in Figure 9. Both artifact on the boundary and RMSE increases with time. Variation of this RMSE with time is shown in Figure 10. It shows that the least error in reconstruction corresponds to 0-20 min scanning duration for every case. From Figure 8(b), it is also shown that if we increase the time window, RMSE increases.

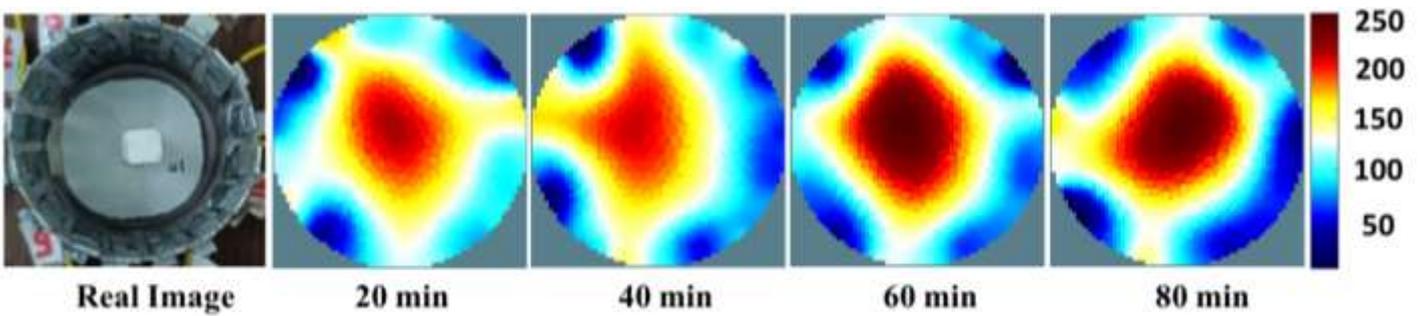

**Figure 9**. Effect of scanning duration on reconstruction

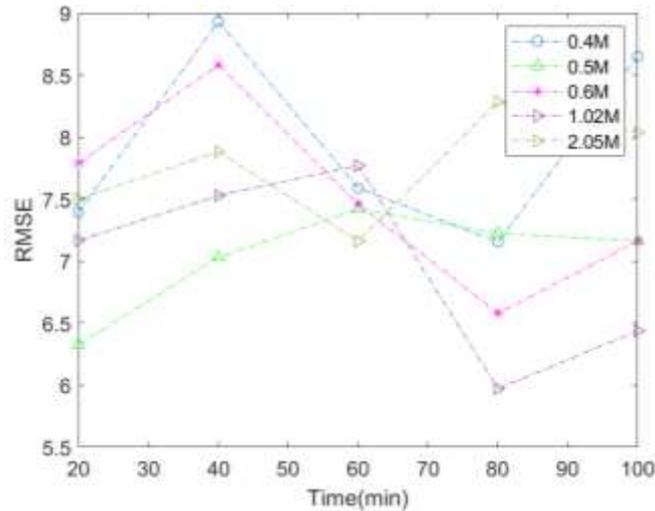

**Figure** 10. Variation of RMSE w.r.t. scanning duration; except 1.02 molar solution all show 20 minute is the optimal scanning duration.

Time may be an important factor if we are considering saline solution as a continuous media. Ions from saline solution migrate towards opposite electrodes when AC current is applied. This salt deposition increases with time. If we have filled the molar solution in a tank or EIT vessel, then the experiment should be performed within 20 minutes. From 20 to 40 minutes, RMSE rapidly increases, which shows rapid deposition of ions on electrodes on excitation. This deposition may increase with time.

Further, the weight of the EIT vessel with electrodes (without saline water filled and without sample) is measured before and after the experiment to find the weight of the deposited salt. It is found that from the first 20 minutes, the weight of the vessel is almost the same, and after 20 min, this weight increase means the deposition of salt on electrodes increases which leads to a decrease in the accuracy of the experimental results. From graph 11(a), there is almost a flat region from 0-20 min; that region is best for EIT experiments. The EIT vessel's weight change with time is shown in Figure 11(a). The condition of the saline tank after the salt deposition is shown in Figure 11(b). Higher scanning duration, for example, 80 minutes, is impractical for most of the applications.

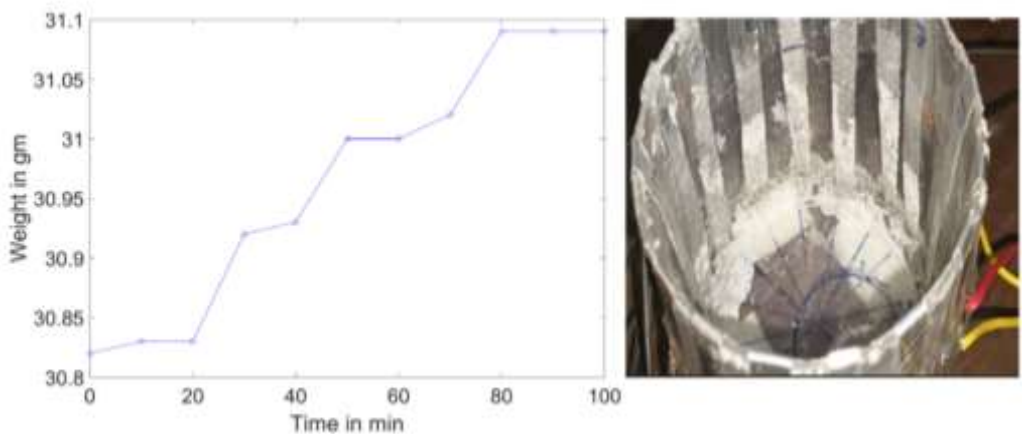

Figure 11(a). Salt deposition

Figure 11(b). Condition of the saline tank after the salt deposition

### 3.2.3 relation to detect material distribution

This study is carried out using variable vessel size, object size, number of electrodes, the width of electrodes, scanning duration, and molarity. PLA samples with variable sizes from 0.2 cm to 1.9 cm are used as material distribution inside. All the PLA samples used in the experiment are shown in Figure 7. The size of the object may be another important parameter of the EIT system. If we have taken an object as large as possible (up to the size of the vessel), then the reconstructions are less noisy. But this is not true for objects with small sizes. We have used five vessels of different sizes. Information of those vessels is given in Table 2. Reconstructions of a few data sets are shown in Figure 12. Using each vessel, we scanned all PLA objects by changing all six design optimization parameters one by one to check their effect.

Empirical expression (shown in Equation 2) is obtained by fitting all experimentally obtained data by varying all six design parameters. Information about non-linear fitting is provided in S3.

Experimentally it is found that if the area of the object ($A_S$), number of electrodes (n), area of the vessel ($A_V$), scanning duration (T), percentage periphery covered by the electrodes (P), and molarity of coupling media (M) follow the proposed relation (equation 2), then we will be able to recover the inner object profile with 10% RMSE error (in worst case scenario).

A random measurement case is tested now. Random values of M, T, n, $A_S$, $A_V$, and P are used to carry out an experiment. RMSE is estimated for the reconstructed image. The same values are fed to equation 2 as well, and the predicted RMSE is estimated. The RMSE between the predicted and experimental values is found to be 1.4053.

$$-0.0007(M)^{0.0083} + 0.0010(T)^2 - 0.038(n)^{-0.0325} + 0.01\left(\frac{A_S * P}{A_V}\right)^{0.011} \leq 10 \qquad ...(2)$$

Table 2. Information of the EIT vessels used

| Name of Vessel | Size of vessel (cm) | Percentage periphery covered by electrodes (%) | Predicted Spatial resolution (mm) | Experimentally Obtained Spatial Resolution(mm) |
|---|---|---|---|---|
| vessel 1 | 5.4 | 38.66 | 0.01529 | 2 |
| vessel 2 | 6.5 | 39.19 | 0.01841 | 2 |
| vessel 3 | 8.3 | 50.88 | 0.02350 | 2 |
| vessel 4 | 19 | 35 | 0.05382 | 4 |
| vessel 5 | 21.7 | 35 | 0.06147 | 7 |

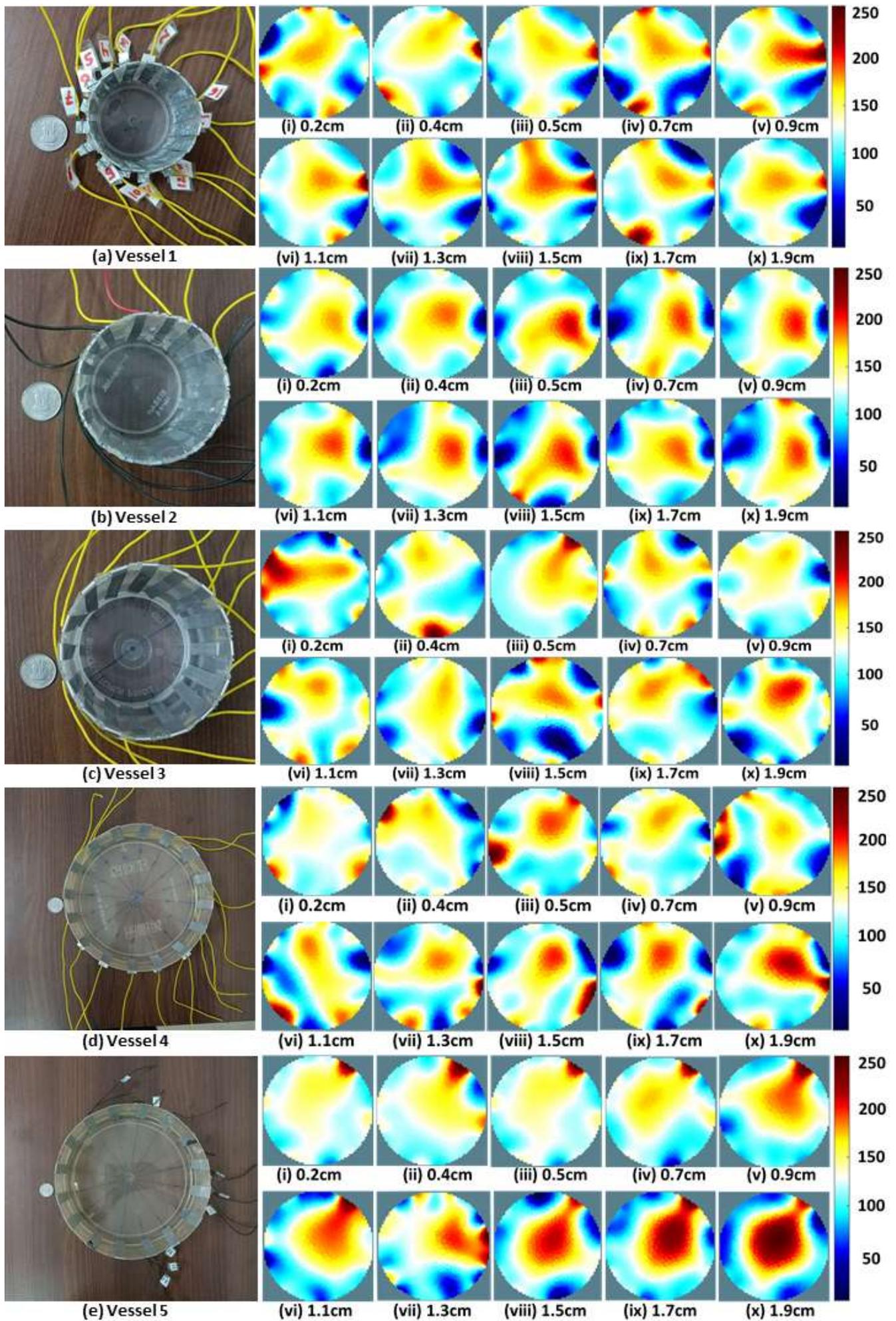

**Figure 12.** Saline tanks　　　　Reconstruction of different size of samples using saline tanks

## Discussion

Equation (2) suggests that all the variables contribute to image reconstruction, but the significant contribution comes from scanning duration. The effects of molarity and number of electrodes relative to scanning duration are weak. For clinical and industrial applications, parameters like molarity, size of the object, and size of the vessel may not be in our control, so one can minimize the RMSE in reconstruction by controlling the scanning duration and periphery of the vessel covered by the electrodes. The maximum value of molarity is 2M for acceptable RMSE using the developed system.

## Conclusion:

EIT systems are developed to study the effect of design and operating parameters. The quality of recovered images is used as a performance parameter.

This study reports that besides previously reported parameters, six other parameters exist that affect the image reconstruction (i) Salinity of coupling media, (ii) Scanning duration, and (iii) Size of the system and the object.

The novelty aspect of this work is the empirically fit expression, including these design and operating parameters to provide a worst-case error in reconstruction.

It is shown qualitatively and quantitatively that the saline solutions' (as a coupling media) molarity affects the image reconstruction. Scanning time also affects the image quality due to ion accumulation on the electrodes. Based on experimental results, we conclude that scanning duration and the molarity of the solution after a particular range can lead to higher RMSE and affect image reconstruction. Also, a relation between EIT setup and object size must be followed before designing a specific EIT setup.

### Acknowledgments:
Ms. Vaishali Sharma is thankful to CSIR, India, for awarding JRF and SRF Fellowship.
### CRediT authorship contribution statement
VS: Methodology, Investigation, Visualization, Software, Instrumentation, funding; MG: Methodology, Investigation, Writing, funding.
### Conflicts of Interest: The authors declare no conflict of interest.
### Availability of data and materials:
The datasets created and/or analyzed during the current study will be made available from the corresponding author upon reasonable request.